\documentclass[runningheads, a4paper]{llncs}

\usepackage{amssymb}
\usepackage{amsmath}
\usepackage[english]{babel}
\usepackage{centernot}
\usepackage[T1]{fontenc}
\usepackage[colorlinks = true,
            linkcolor = blue,
            urlcolor  = blue,
            citecolor = blue,
            anchorcolor = blue]{hyperref}
\usepackage[utf8]{inputenc}
\usepackage{lmodern}
\usepackage{listingsutf8} 
\usepackage{lstcoq}
\usepackage{mathtools}
\usepackage{nameref}
\usepackage{underscore}
\usepackage{xspace}

\newcommand{\compl}[1]{\overline{#1}}
\newcommand{\compose}{\otimes}
\newcommand{\dl}{\(\mathbf{d}\mathcal{L}\)\xspace}
\newcommand{\E}{\mathcal{E}}
\newcommand{\glb}{\sqcap}
\newcommand{\hcsp}{\textsc{hcsp}\xspace}
\newcommand{\csp}{\textsc{csp}\xspace}

\newcommand{\implements}{\vdash}
\newcommand{\inter}{\cap}
\newcommand{\M}{\mathcal{M}}
\newcommand{\mars}{M\textsc{ars}\xspace}
\newcommand{\nin}{\centernot\in}
\newcommand{\provides}{\vdash_E}
\newcommand{\refines}{\preccurlyeq}

\DeclareMathOperator{\saturate}{saturate}

\newcommand{\union}{\cup}
\title{A Mechanically Verified Theory of Contracts}
\titlerunning{A Mechanically Verified Theory of Contracts}

\author{Stéphane~Kastenbaum\inst{1,2} \and
Benoît~Boyer\inst{2} \and
Jean-Pierre~Talpin\inst{1}}

\authorrunning{S. Kastenbaum et al.}

\institute{Inria Rennes - Bretagne Atlantique, France
\and
Mitsubishi Electric R\&D Centre Europe, Rennes, France
}

\begin{document}

\maketitle

\begin{abstract}
  Cyber-physical systems (CPS) are assemblies of networked, heterogeneous, hardware, and software components sensing, evaluating, and actuating a physical environment.
This heterogeneity induces complexity that makes CPSs challenging to model correctly.
Since CPSs often have critical functions, it is however of utmost importance to formally verify them in order to provide the highest guarantees of safety.
Faced with CPS complexity, model abstraction becomes paramount to make verification attainable. 
To this end, assume/guarantee contracts enable component model abstraction to support a sound, structured, and modular verification process.
While abstractions of models by contracts are usually proved sound, none of the related contract frameworks  themselves have, to the best of our knowledge, been formally proved correct so far.
In this aim, we present the formalization of a generic assume/guarantee contract theory in the proof assistant Coq.
We identify and prove theorems that ensure its correctness.
Our theory is generic, or parametric, in that it can be instantiated and used with any given logic, in particular hybrid logics, in which highly complex cyber-physical systems can uniformly be described.
\end{abstract}

\section{Introduction}

With the rise of cyber-physical systems with growingly critical functions, it becomes of the utmost importance to develop frameworks that support their sound design and guarantee their safety.
Currently, to design such systems, engineers use informally specified tools such as Simulink~\cite{dabney_mastering_2003}.
Simulink is a very convenient toolbox to connect components to form systems.
From atomic components, systems become components in larger systems that can be reused in multiple places in a given design.
As an effort to verify Simulink systems, formal verification frameworks such as \mars translate Simulink design into a formal specification language~\cite{chen_mars_2017,liebrenz_deductive_2018}.
This approach alters the native hierarchy of Simulink models in the specification but allows modularity to be reconstructed afterward in the target formal language.
Ideally, however, we would like to conserve the intended component hierarchy for the verification process, in order to better support abstraction in places of the design.
Contracts are a tool to support the abstraction of subsystems in such specification formalisms.

Contracts help to design and to verify complex systems by abstracting component models, using
their assumptions and guarantees in place of their exact, internal specification.
A contract being more abstract than the specification of a component makes the modular verification of large systems feasible.

In this paper, we first review the related works.
We then give an overview of the contract theory proposed by Benveniste~et~al.~\cite{benveniste_contracts_2015,benveniste_multiple_2008}.
The subsequent section presents the formalization of our theory in the proof assistant Coq.
We then illustrate its instantiation with a simple propositional logic.
This gives us the opportunity to discuss some design choices in the formalization.
Finally, we outline future works and conclude.

\section{Related Works}
\label{related-works}

A recent interest in creating contract frameworks for cyber-physical systems is appearing.
The complexity of such systems and their interconnections make the verification and validation process challenging, render traditional system design methods inadequate, and call for more powerful frameworks to design such complex systems~\cite{graf_building_2018,sangiovanni-vincentelli_taming_2012}.

Design by contracts was first proposed by Meyer for software programming~\cite{meyer_applying_1992}.
Hence, specification by contracts traditionally consists of pre- and post-conditions, leaving the continuous timed variables extraneous of the specification.
This is not practical when designing cyber-physical systems, where time is intrinsically linked to the continuous behavior of the system.
In contracts for cyber-physical systems, we replace pre-conditions by assumption and post-condition by guarantee, the main difference being that the assumption and guarantee can express properties of continuous timed variables.

To design cyber-physical systems, numerous hybrid logics have been proposed.
Differential dynamic logic (commonly abbreviated as \dl) defines hybrid programs, a combination of discrete computation and differential equations to model cyber-physical systems~\cite{platzer_differential_2008}.
It is equipped with a proof assistant, Keymaera~X, to check and prove safety and reachability properties on the modeled systems~\cite{fulton_keymaera_2015}.
While the verification process supports decomposition, it does not support parallelism hence modularity.
Multiple approaches have therefore been investigated to define contracts in \dl~\cite{lunel_parallel_2019,muller_tactical_2018}.
Their goal is to define a composition theorem allowing to connect multiple components.
Both approaches conceptualize components as a design pattern on hybrid programs and design contracts as abstract specifications around them.

Hybrid \csp is an extension of \csp with differential equations used to model cyber-physical systems~\cite{chaochen_formal_1996}.
A language to describe the trace of executions, the duration calculus, is used in conjunction with \hcsp to verify properties of cyber-physical systems~\cite{chaochen_extended_1993,wang_improved_2015}.
This is done in the proof assistant Isabelle~\cite{nipkow_isabellehol_2002}.
Since \csp supports composition, work to define contracts in \hcsp have focused on the composition of abstract specifications in the duration calculus~\cite{wang_assumeguarantee_2012}.

Other approaches have been considered to define contracts for cyber-physical systems.
For example, contracts defined with Signal Temporal Logic were used to ensure the safety of autonomous vehicles in~\cite{arechiga_specifying_2019}.

While slightly different from one another, all these definitions of contracts have the same core ideas.
Namely, a contract abstracts the specification of a component in the same logic.
These definitions also support the same usual operators on contracts such as composition or refinement.
A meta-theory of contracts has been defined, aiming at unifying all theories of contracts~\cite{benveniste_contracts_2015}.
The theory of assume/guarantee contract instantiates the meta-theory and is generic enough to bridge the gap between related definitions of contracts~\cite{abadi_composing_1993,benveniste_multiple_2008}.
It was used, for instance, to define contracts in heterogeneous logics and relate them together~\cite{nuzzo_compositional_2015}.
It is given an overview in Section \ref{theory}.

Foster et al. have proposed a mechanized theory of contracts in Isabelle and the Unified Theory of
Programming (UTP) using pre-, peri- and post-conditions to abstract discretely timed systems.~\cite{foster_unifying_2019}

Yet the meta-theory of contracts and the theory assume/guarantee of contracts have, to the best of our knowledge, no formalized proof of correctness.
In the following sections, we propose a formalization of both of them in the proof assistant Coq with the goal to instantiate them with a hybrid logic that could be defined using that theorem prover~\cite{the_coq_development_team_coq_2018}.

\section{Overview of the Meta-Theory and the Assume/Guarantee Theory of Contract}
\label{theory}
In this section, we recall the meta-theory of contracts introduced by Benveniste et~al.~\cite{benveniste_contracts_2015}.
Then the definition of one of its implementation: the set-theoretic assume/guarantee contract theory~\cite{benveniste_multiple_2008}.

  \subsection{Meta-Theory}

In the meta-theory of contracts, there is only one notion, that of component.
That notion is kept abstract, it is not defined.
In practice, it is meant to represent an element of a system performing a specific task.
In the remainder, components are noted by the letter $\sigma$.

Components are subject to a relation of compatibility which, in the meta-theory, is kept abstract as well.
Two compatible components can be composed to create a larger component or subsystem.
Composition is akin to connecting two elements of a system together.
We note $\sigma_1 \times \sigma_2$ for the composition of components $\sigma_1$ and $\sigma_2$.

We call \emph{environment} of $\sigma$ any component that is compatible with $\sigma$.
For example, when considering the motor of a car, the other components interacting with the motor form a sub-system: the environment of the motor.
Conversely, when considering the gearbox, the other components of the car, including the motor, are the environment.

A contract is the specification of a task in a system.
We model a contract with a pair $C = (\E, \M)$ with $\M$ a set of components and $\E$ a set of environments compatible with every component in $\M$.
Multiple components can perform that same task, all must respect its contract.
We say that a component $\sigma$ \emph{implements} the contract $C$, if $\sigma \in \M$.
We use the notation $\sigma \implements C$ for $\sigma$ implements $C$.
A contract is also  a specification of the environment.
It describes any environment in which the task can be achieved by the components.
Dually, we note $e \provides C$ to say that $e$ \emph{provides} the contracts, meaning $e \in \E$

The goal of a specification is to be an abstraction on actual components.
Multiple contracts can be the specifications of the same component on different levels of abstraction.
We introduce the \emph{refinement} relation to describe that a contract is the abstraction of another.
Here $C_1$ is the refined version of $C_2$.
This means that any implementations of $C_1$ can be used in place of an implementation of $C_2$.
\begin{definition}[Refinement]
\label{mt-refinement}
  $C_1 \refines C_2 \equiv \M_1 \subseteq \M_2 \land \E_2 \subseteq \E_1$
\end{definition}

Sometimes, multiple specifications can be applied to the same component.
For example, we want a component to be fast and correct.
There is both a specification for the speed of the component and one for the correctness.
Therefore, we want to regroup both specifications on the same contract.
Any contract which refines both contracts can be used, though it is more desirable to use the most abstract contract.
With this in mind, we define the conjunction of contracts as the greatest lower bound of refinement.
\begin{definition}[Conjunction]
\label{mt-glb}
  $C_1 \glb C_2$ is the greatest lower bound of contract on refinement.
\end{definition}

For now, we only have considered one component in its environment.
Yet most of the complexity of systems comes from the composition of multiple components.
The problem can be formulated as \emph{"If we have the specification of two components, can we determine the specification of the composition of the components?"}.
The subtlety is that each component is part of the environment for the other.
So the result of their specifications takes into account this point.
\begin{definition}[Composition]
\label{mt-compositon}
  $$
  C_1 \compose C_2 \equiv \min \left\{C\ {\Bigg|}\ 
  \begin{array}{lcl}
  \forall M_1 \implements C_1&           & M_1 \times M_2 \implements C  \\
  \forall M_2 \implements C_2& \implies  & E \times M_2 \implements C_1 \\
  \forall E \provides C &           & E \times M_1 \implements C_2
  \end{array}
  \right \}
  $$
\end{definition}

The definition of quotient, lowest upper bound, compatibility, and consistency of contracts are not relevant for our purposes.
The interested reader is directed toward their definitions in~\cite{benveniste_contracts_2015}.

As much as this meta-theory gives us, it is not enough to be used.
It only allows us to understand what are the necessary tasks needed to define a contract theory.
The meta-theory is not constructive, for example, the $\min$ operator used in the definition of composition doesn't give us the actual contract.
In the next section, we define an actually usable contract theory, without losing any generality.

  \subsection{Assume/Guarantee Contract Theory}
\label{ag-theory}

This section defines a contract theory by instantiating the meta-theory with concrete definitions, starting with that of a component.
We model a component by the properties it guarantees on the system.

In this state-centered theory, there is no clear demarcation between inputs and outputs.
We note $d$ the set of all variables.
For simplification purposes, all variables hold value the same domain $B$.

A state is a valuation of all the variables, that is to say, it is a surjective function from $d$ to $B$.
The state space $S : d \to B$ is every configuration the system can be in.

In system design, an assertion is a property of the state the system is in.
We use the duality of sets, it can be seen either as a collection of elements or as a property that every element satisfies.
Here, we consider the assertion as every state that satisfies the property.
This means, we define an assertion $A$ as a subset of the state space, noted $A \subseteq S$.
A component is not only viewed as a property it ensures on the variables but as every state satisfying this property.
Which means a component is a set of states.

Here every component is compatible with every other component.
Meaning every component can be composed, or connected, to any other component.
The composition of two components is simply the intersection of their assertions.
This means that every component can be seen as an environment for another.
We should mention that if a component is an empty set, it is a non-implementable component.

Using contracts as defined in the meta-theory above is not practical, we use another definition and will prove later that they're equivalent.
Here, contracts are associations of assumption and guarantee.
The assumption is the input accepted by the component and the guarantee is properties ensured on the output of the component.

A contract is an abstract specification of a component.
The goal is to define rules which restrain both the implementation of the component and the environment it needs to be in.
Here, the assumption is the restriction on the environment whereas the guarantee is the restriction on the component.
\begin{definition}[Assume/guarantee contract]
  An assume/guarantee contract is the combination of two assertions, one for the assumption ($A$) one for the guarantee ($G$).
  For a contract $c = (A,G)$ we define projections to get the assumption and the guarantee
  $$A(c) = A\ ;\ G(c) = G$$.
\end{definition}

Now, we define what it means for a state to \emph{satisfy} a contract.
A state satisfies a contract if either it's a state excluded from the contract's assumption, or the contract's guarantee holds for the state.
\begin{definition}[Satisfies]
  $$ s \vdash c \equiv s \in \compl{A(c)} \cup G(c) $$
\end{definition}

We can lift this definition to components.
A component $\sigma$ implements a contract if every state in $\sigma$ satisfies the contract.
\begin{definition}[Implements]
  $$ \sigma \implements c \equiv \forall s \in \sigma,\ s \vdash c$$
\end{definition}

A environment $e$ provides a contract if every state of $e$ is included in the assumption of the contract.
\begin{definition}[Provides]
    $$ e \provides c \equiv \forall s \in e,\;s \in A(c)$$
\end{definition}

This leads to a particular point in assume/guarantee contracts.
With the above definition, we notice that multiple contracts can be implemented by the same components and provided by the same environments.
So we have a class of contracts which are all equivalent, as they specify the same set of components and environment.
We define \emph{saturation} an idempotent operation which doesn't change the set of components satisfying the contract, nor the set of environment providing the contract.
We then always use the saturated version of a contract.
\begin{definition}[Saturation]
  $$\saturate(c) \equiv (A(c),\;\compl{A(c)} \cup G(c))$$
\end{definition}

Next, we want to define the refinement relations between two contracts.
There is a simple definition of refinement.
The most refined contract needs to have stronger guarantee, and looser assumption.
In the formalization, we prove that it is equivalent to  Definition~\ref{mt-refinement}.
\begin{definition}[A/G refinement]
\label{ag-refinement}
  $$c1 \refines c2 \equiv A(c1) \supseteq A(c2) \land G(c1) \subseteq G(c2)$$
\end{definition}
This definition is decidable if and only if $\subseteq$ is decidable.
We do want the refinement relation to be decidable, hence we have to make sure that $\subseteq$ is decidable.

To find the conjunction of contracts, (or the greatest lower bound of refinement), we have the formula below.
Contrary to Definition~\ref{mt-glb} in the meta-theory, this definition is constructive.
\begin{definition}[A/G conjunction]
\label{ag-glb}
  $$ c1 \glb c2 \equiv (A(c1) \cap A(c2), G(c1) \cup G(c2)) $$
\end{definition}

Composition of contracts is defined as followed, this definition is constructive.
It also is the $\min$ in the sens of refinement as expressed in Definition~\ref{mt-compositon}.
\begin{definition}[A/G composition]
\label{ag-composition}
  $$c1 \compose c2 \equiv (A', G')$$
  With $$A' \equiv A(c1) \cap A(c2) \cup \compl{G(c1) \cap G(c2)}$$
  $$G' \equiv G(c1) \cap G(c2)$$
\end{definition}

We now have two definitions of contracts.
The key difference is the definition of refinement, conjunction, and composition.
The meta-theory has an intuitive definition, while the assume/guarantee theory has more interesting properties.
In the next section, we formalize the assume/guarantee theory of contract and prove that it is equivalent to the above meta-theory of contract.

\section{Formalization of Assumption/Guarantee Contract Theory}
\label{formalization}

The meta-theory intends to provide a generic contract theory that can be instantiated by several logics.
Each logic presents some features that enable or facilitate the verification of system properties.
Proving that several logics implement the same meta-theory is a way to unify them.
Here, we formalize the assume/guarantee contract theory and prove it to correspond to the definitions given in the meta-theory.

The contract theory relies on set-theoretic definitions.
At this stage, we assume the abstract type \lstinline{set : Type -> Type} which is equipped with the usual set operators as $\cup$, $\cap$, $\neg$.
We also assume the relations $\in$ and $\subseteq$ as well as the set equivalence \lstinline|s1 == s2| which is extended to the standard equality \lstinline|s1 = s2| by extensionality.

In this section, we give an overview of our formalization of the contract theory.
First, we consider every component to be defined on the same variables, next, we see how to handle multiple variable sets.

  \subsection{Single Variable Set}
For a given set of variables, the semantics of a contract is defined by the assertion of its assumption and guarantee, each represented by a set of states.
A state is a valuation from variables to values, where the variables are a set of identifiers as in the following:

\begin{lstlisting}
Variable vars : set ident.
Definition var := { v : ident | v $\in$ vars }.
Definition state := var -> B.
Definition assertion := set state.
Definition component := assertion.
Definition environment := component.
Record contract : Type := mkContract { A : assertion ; G : assertion }.
\end{lstlisting}
The types \lstinline|ident| and \lstinline|value| are parameters of our theory and kept abstract, but we require that identifiers must be discriminable and their membership to sets must be decided.
\lstinline{vars} is the set of identifiers used in the system, and \lstinline{var} is the type of a variable, namely an identifier with the proof that it's in \lstinline{vars}.
At this stage of the development, we assimilate the concept of a component with its behavior as in Section~\ref{ag-theory}.

We define assertions as state predicates using the duality of sets, namely, \lstinline|s $\in$ q| denotes that \lstinline|s| satisfies the assertion \lstinline|q|.
Contracts are directly defined as pairs of assertions relating the behavior expected from the environment (assumption) with the behavior of the component (guarantee).
The syntax \lstinline|c.A| (and \lstinline|c.G|) denotes the assumption (respectively guarantee)  of the contract \lstinline|c| in the rest of the paper\footnote{\footnotesize The Coq's original syntax is  \lstinline|c.(A)| but we replaced it for the sake of readability.}.

The semantics of the contract relies on the implementation of a contract by a component.
In order to define it, we first introduce the satisfiability of the contract by a single state and the saturation principle.

\begin{lstlisting}
Definition satisfies (s : state) (c : contract) : Prop :=
    s $\in$ $\neg$ c.A $\cup$ c.G.
\end{lstlisting}
Basically, a state satisfies a contract either if the state is discarded by the assumption and nothing is guaranteed by the contract, or the state satisfies both assumption and guarantee of the contract.

In the following code, we saturate contracts when necessary, indeed it's easier to saturate a contract than to check if it's already saturated.
Whereas in the mathematical definition it is easier to consider every contract to be saturated than to saturate every time.
In the code, we \lstinline{saturate} contracts when needed, whereas in the mathematical definition we have considered contracts to be saturated.

\begin{lstlisting}
Definition saturate (c : contract) : contract :=
    {| A := c.A ;
       G := $\neg$ c.A $\cup$  c.G |}.
\end{lstlisting}

The contract saturation is sound: the same states are characterized before and after the saturation of any contract.
\begin{lstlisting}
Theorem saturate_sound : forall (s : state) (c : contract),
    satisfies s c <-> satisfies s (saturate c).
\end{lstlisting}

We extend the contract satisfiability from states to components to define the implementation relation.
Additionally, we also need to characterize the relationship between contract and environment.

\begin{lstlisting}
Definition implements ($\sigma$ : component) (c : contract) :  Prop :=
    forall s, s $\in$ $\sigma$ -> satisfies s c.
Notation "$\sigma$ $\implements$ c" := (implements $\sigma$ c).
Definition provides (e : environment) (c : contract) : Prop :=
    e $\subseteq$ c.A .
\end{lstlisting}

Then, we can define the refine relation on contracts.
Here, it is important to note that we are implementing the assume/guarantee theory of contracts.
The refinement and composition relation are defined differently in the meta-theory and in the assume/guarantee theory.
\begin{lstlisting}
Definition refines (c1 c2 : contract) : Prop :=
    let (c1' , c2') := (saturate c1 , saturate c2) in
    c2'.A $\subseteq$ c1'.G /\ c1'.G $\subseteq$ c2'.G.
Notation "c1 $\preceq$ c2" := (refines c1 c2).
\end{lstlisting}
Since \lstinline|refines| is an order, we proved the usual properties: reflexivity, transitivity, and antisymmetry.
We also demonstrated that this set-theoretic definition is equivalent to the more standard and meaningful Definition~\ref{mt-refinement} of the refinement given by the meta-theory.
\begin{lstlisting}
Theorem refines_correct : forall (c1 c2 : contract),
    c1 $\refines$ c2 <->
    (forall $\sigma$: component, $\sigma$ $\vdash$ c1 -> $\sigma$ $\vdash$ c2) /\
    (forall e: environment, provides e c2 -> provides e c1).
\end{lstlisting}

The conjunction of contracts corresponds to the multiple views, one can have on the same component.
In the meta-theory, it is defined as the greatest lower bound of refinement.
So, we prove that our set definition is equivalent to the meta-theoretical Definition~\ref{mt-glb}.
\begin{lstlisting}
Definition glb (c1 : contract) (c2 : contract) : contract :=
    let c1' := saturate c1 in let c2' := saturate c2 in
    mkContract (c1'.A $\cup$ c2'.A) (c1'.G $\cap$ c2'.G).
Notation "c1 $\glb$ c2" := (glb c1 c2). 

Theorem glb_correct : forall c1 c2 : contract,
    (c1 $\glb$ c2) $\refines$ c1 /\ (c1 $\glb$ c2) $\refines$ c2 /\
    (forall c, c $\refines$ c1 -> c $\refines$ c2 -> c $\refines$ (c1 $\glb$ c2)).
\end{lstlisting}

The central operator in contracts algebra is the composition.
Two contracts can be composed if they are defined on the same variables.
The composition of components aims to construct a contract specifying the composition of components.
We provide a set-theoretic definition and then show that it corresponds to the meta-theory Definition~\ref{mt-compositon}. 

\begin{lstlisting}
Definition compose (c1 c2 : contract) : contract :=
    let c1' := saturate c1 in
    let c2' := saturate c2 in
    let g := c1'.G $\cap$ c2'.G in
    let a := (c1'.A $\cap$ c2'.A) $\cup$ $\neg$ g in
    mkContract a g.
Notation "c1 $\compose$ c2" := (compose c1 c2).
\end{lstlisting}

Here again, we give the proof that it corresponds to Definition~\ref{mt-compositon} given in the meta-theory.

\begin{lstlisting}
Theorem compose_correct :
    forall (c1 c2 : contract) ($\sigma$1 $\sigma$2 : component) (e : environment),
    $\sigma$1 $\implements$ c1 -> $\sigma$2 $\implements$ c2 -> provides e (c1 $\compose$ c2) ->
    ($\sigma$1 $\inter$ $\sigma$2 $\implements$ c1 $\compose$ c2 /\ provides (e $\inter$ $\sigma$2) c1 /\ provides (e $\inter$ $\sigma$1) c2).
  
Theorem compose_lowest : forall (c1 c2 c : contract), 
    (forall ($\sigma$1 $\sigma$2 : component) (e : environment),
    $\sigma$1 $\implements$ c1 -> $\sigma$2 $\implements$ c2 -> provides e c ->
    ($\sigma$1 $\inter$ $\sigma$2 $\implements$ c /\ provides (e $\inter$ $\sigma$2) c1 /\ provides (e $\inter$ $\sigma$1) c2)) ->
    c1 $\compose$ c2 $\refines$ c.
\end{lstlisting}

The above definitions all consider every state to be defined on the same variables.
Yet, it is highly improbable that every component needs every variable to be defined.
Thus, some components are defined on different sets of variables.
To define the composition of components defined on different sets of variables, we need to extend their definitions to larger sets.
We may also want to eliminate variables out of contracts.
If a component allows for every possible value of a variable and doesn't give any guarantee on it there is no need to have it in the specification.

  \subsection{Alphabet Equalization}

This section considers the case when contracts are defined over different domains of variables.
Two problems can occur in this situation.
First, we may need to compose two contracts that are not defined on the same variables.
In that case, we need a way to extend the two contracts on the union of their variables.
Then, some variables in a contract may be useless.
For example, if a component provides the input for another component, maybe the composition of the two contracts specifying the components doesn't need to specify this variable.
In that case, we need the elimination of variables.

In \cite{benveniste_multiple_2008}, the authors defined elimination of variables and extension of contracts.
The elimination of variables is defined in the following way.
\begin{definition}
\label{elimvar}
For a contract $c = (A,G)$, and a variable $v$ in the contract.
The contract without $v$ is :
  \[ [c]_v \equiv ( \forall v\  A \ ;\ \exists v \ G) \]
\end{definition}
This definition is not usable in our formalization.
$A$ and $G$ are both sets, the quantifier has no meaning for sets.
To define this elimination of variable, the authors consider assertions as logic formulas and bind free variables with quantifiers.
However, this shortcut cannot be taken with a formal proof assistant.
The definition of the extension of contracts was also eluded by considering assertion as logic formulas.

In the following, we give a set-theoretic definition in Coq of the elimination of variables in a contract and the definition of the extension of contracts.
We assume \lstinline{d1} and \lstinline{d2} two variable sets, with \lstinline{H12 : d1 $\subseteq$ d2}.

First, we define \lstinline{H'12 : var d1 -> var d2}, which takes a variable in d1 and shows that it is also a variable in d2.
\begin{lstlisting}
Definition H'12 (v1 : var d1) : var d2 := let (i,H1) := v1 in exist _ i (H12 i H1).
\end{lstlisting}
Here, we use \lstinline{H12} to show that the ident \lstinline{i} in \lstinline{d1} is also in \lstinline{d2}.
Indeed, the type of \lstinline{H12} is \lstinline{forall v : ident, v $\in$ d1 -> v $\in$ d2}.
Hence, \lstinline{H12 i H1} is of type \lstinline{i $\in$ d2}, and \lstinline{exist _ v (H12 i H1)} is of type \lstinline{var d2}.

We define the projection of a single state on a smaller variable set.
\begin{lstlisting}
Definition project (e2 : state d2) : state d1 :=
    fun v1 => e2 (H'12 v1).
\end{lstlisting}

We extend the definition of projection of state to the projection of assertion.
Which is the projection of every state in it.
If we consider the assertion as the property it holds on variables, the projection is similar to the existential quantifier.
For example, the assertion $P(x,y)$ projected on the variables $\{y\}$ is $\exists x, P(x,y)$.
\begin{lstlisting}
Definition project_assertion (a : assertion d2) : assertion d1 :=
    fun e1 => exists e2,  e2 $\in$ a /\ project e2 = e1.
\end{lstlisting}

Then, we can define the inverse of the projection.
Which for a state gives the set of states that project to it.
Notice that the extension of a state gives an assertion.
Multiple states defined on \lstinline{d2} have the same projection on \lstinline{d1}.
\begin{lstlisting}
Definition extend_state (e : state d1) : assertion d2 :=
    fun e2 => project e2 = e.
\end{lstlisting}

Similar to the extension of states, extending an assertion \lstinline{a1} is done by taking every state that projects to a state in \lstinline{a1}.
\begin{lstlisting}
Definition extend_assertion (a1 : assertion d1) : assertion d2 :=
    fun e2 => project e2 $\in$ a1.
\end{lstlisting}

Next, we need to define the \emph{strong projection} of an assertion.
Which is the set of states where every extension of the states are in the assertion.
The equivalent of the strong projection when viewing assertion as property is the $\forall$ quantifier.
The strong projection of $P(x,y)$ on $\{y\}$ is $\forall x,\, P(x, y)$.
\begin{lstlisting}
Definition project_assertion_forall (a : assertion d2) : assertion d1 :=
    fun e1 => extend_state e1 $\subseteq$ a.
\end{lstlisting}

Finally, we can define the elimination of variables in a contract.
Eliminating variables is the same thing as projecting onto the other variables.
We reuse Definition~\ref{elimvar} with the set definitions of projection.
\begin{lstlisting}
Definition project_contract (c2 : contract d2)  : contract d1 :=
    let c2' := saturate _ c2 in
    mkContract _ (project_assertion_forall c2'.A)
                  (project_assertion c2'.G).
\end{lstlisting}

Extending a contract onto a bigger variable set is extending both the assumption and the guarantee.
In the set theory, this part is implicit.
With the proof assistant, we have to make it explicit.
\begin{lstlisting}
Definition extend_contract (c1 : contract d1) : contract d2 :=
    let c1' := saturate _ c1 in
    mkContract _ (extend_assertion (c1'.A) (extend_assertion (c1'.G)).
\end{lstlisting}

We can now define composition on two contracts defined on different variables.
We assume \lstinline{d1} \lstinline{d2} and \lstinline{d3}, with \lstinline{H1 : d1 $\subseteq$ d2} and \lstinline{H2 : d2 $\subseteq$ d3}.
\begin{lstlisting}
Definition extended_compose (c1 : contract d1) (c2 : contract d2) :
    contract d3 :=
    compose _ (extend_contract H1 c1) (extend_contract H2 c2).
\end{lstlisting}

We can partially verify the correction of the extended composition, though we still need to find a satisfying correction theorem.
For example, we verify that the composition of contracts implements the composition of components on different sets of variables.
\begin{lstlisting}
Theorem extended_compose_correct : forall (c1 : contract d1) (c2 : contract d2)
    ($\sigma$1 : component d1) ($\sigma$2 : component d2),
    implements _ $\sigma$1 c1 -> implements _ $\sigma$2 c2 ->
    implements _ (extend_assertion H1 $\sigma$1 $\union$ extend_assertion H2 $\sigma$2)
        (extended_compose c1 c2).
\end{lstlisting}

\section{Instantiating the Assume/Guarantee Contract Theory}

Our goal is to have a formally verified generic theory for contracts.
With this generic theory, one could define contracts with any propositional logic.
The requirement to use the contracts theory is to give the alphabet of variables and the set of values.
Three properties are also required: the set of values need to be inhabited, and both equality and the $\in$
relation must be decidable.

In a nutshell, one needs to instantiate this interface :
\begin{lstlisting}
Class Theory := {
    B : Type ;
    ident : Type ;
    any_B : B ;
    eq_dec_ident : forall x y : ident, {x = y} + {x <> y} ;
    in_dec_ident : forall (v : ident) (d : set ident), {v $\in$ d} + {v $\nin$ d} ;
}.
\end{lstlisting}

Given these elements, the theory can be instantiated.
The parameter \lstinline{ident} is the type of identifier of the variables.
The parameter \lstinline{B} is the type of values the variables can yield.
We need \lstinline{B} to be inhabited, giving any value with \lstinline{any_B} is sufficient to verify it.
We require equality and the $\in$ relation to be decidable for identifiers.
Hence \lstinline{eq_dec_ident}, \lstinline{in_dec_ident} are required.
This class is used to define states, on which assertion, components, and contracts are defined.

When instantiating the class, the definitions provided suffice to define contracts and components for a given logic.
Yet, defining contracts this way may be too tedious to be practical.
Usually, a logic has a grammar to define formulas, and a "satisfy" relation saying if a state satisfies a formula.
So the workflow is to define the assertion in the language of the logic, then create the set of states using the assertion and the satisfy function.
We aim to simplify this workflow.

An idea is to define contracts and components in the logic, and then show their equivalence with the contracts defined in the meta-theory.
The following example shows how to proceed.

\subsection{Example with a Simple Propositional Logic}

To demonstrate how this theory could be used to create a framework, we developed a simple propositional logic.
We use \lstinline{Prop} as the value type, and \lstinline{nat} as the identifier of variables.
We instantiate the class as the following:
\begin{lstlisting}
Instance theo : Theory := {
    ident := nat ;
    B := Prop ;
    default_B := True ;
    eq_dec_ident := eq_dec_nat ;
    in_dec_ident := in_dec_nat ;
}.
\end{lstlisting}
Where \lstinline{eq_dec_nat} and \lstinline{in_dec_nat} are assumed.
For this instance, we define our states.
\begin{lstlisting}
Definition state (d : set ident) : Type :=  forall x : variable d, B.
\end{lstlisting}

Then we define a formula algebra, with standard connectors and natural numbers to identify variables.
Notice that the formula type is dependent of the variables it is defined on.
We need this to define states properly.
\begin{lstlisting}
Inductive expr {d : set ident} : Type :=
    | f_tt : expr
    | f_var : var d -> expr
    | f_not : expr -> expr
    | f_and : expr -> expr -> expr.
  
Inductive sat {d : set ident} : state d -> expr d  -> Prop := $\dots$
\end{lstlisting}

\noindent To define a contract with two formulas \lstinline{a} and \lstinline{g} as assumption and guarantee, we do:
\begin{lstlisting}
Variable d : set ident.
Definition formula_to_assert (formula : expr) : assertion d :=
    fun e => sat e formula.
Definition mkContractF (a g : expr d): contract d :=
    mkContract d (formula_to_assert a) (formula_to_assert g).
\end{lstlisting}

  \subsection{Logic Specific Contracts}

We may want to define contracts that are only defined in the logic, with a formula as assumption and guarantee.
\begin{lstlisting}
Record contractF :=
    ContractF {A : expr d ; G : expr d}.
\end{lstlisting}

We can define \lstinline{refinesF}, \lstinline{composeF} and \lstinline{glbF}, which are the same as the set definitions, but we replace $\union$, $\inter$ and $\neg$ by \lstinline{f_or, f_and} and \lstinline{f_not}.
With the translation of logic contract to contract in set theory, we can check that the definition of the operators are correct.
\begin{lstlisting}
Definition c2c (cf : contractF) : contract d :=
    mkContractF d (cf.A) (cf.G).
Theorem refinesF_correct : forall (cf1 cf2 : contractF),
    refines d (c2c cf1) (c2c cf2) <-> refinesF cf1 cf2.   
Theorem composeF_correct : forall (cf1 cf2 : contractF),
    c2c (composeF cf1 cf2) == compose d (c2c cf1) (c2c cf2).
Theorem glbF_correct : forall (cf1 cf2 : contractF),
    c2c (glbF cf1 cf2) == glb _ (c2c cf1) (c2c cf2).
\end{lstlisting}
Here, \lstinline{==} is the equivalence of contracts, defined by:
\begin{lstlisting}
Definition equiv (c1 : contract) (c2 : contract) : Prop :=
    refines c1 c2 /\ refines c2 c1.
Notation "c1 == c2" := (equiv c1 c2).
\end{lstlisting}

We have now verified that our contracts are correctly instantiating the theory.
We could use them to design a system and verify properties on it.

\section{Discussion}

In this section, we justify certain design choices.
We explored different ways to define the theory, which we found inefficient for the reasons described below.

  \subsection{State as Function vs. State as Vector}

A prior version of the formalization defined states as vectors of values.
States were defined on a vector of variables, and the index allowed to determine which values corresponded to each variable.
\begin{lstlisting}
Variable n : nat.
Definition state := Vector.t B n.
Definition vars := Vector.t ident n.
Definition assertion := sets state.
Record contract vars := Contract {
    A : assertion ;
    G : assertion ;
}.
\end{lstlisting}

The practical problem of this solution is that it is very difficult to add variables to a state.
Vectors embed their length in the type, which makes composing two contracts defined on different variables impossible .

If we have a contract \lstinline{c1} defined on the variables \lstinline{d1}, and \lstinline{c2} defined on the variables \lstinline{d2}, the composed contract is defined on variables \lstinline{d3 = d1 $\cup$ d2}.
The size of \lstinline{d3} needs to be known before constructing the contract.
But it is not possible, because we don't know the size of \lstinline{d1 $\inter$ d2}.

Even though defining states as vector works on paper, it rises many problems when using a proof assistant.
In definitive, functions are sufficient to model states.
This is why we define states as functions in our formalization.

  \subsection{Variable Set as Type Parameter vs. Variable Set as Record Field}

In our definitions, every state, assertion, or contract depends on the variables it is defined on.
This is a situation where we use the power of dependent type theory.
Another solution would have been to hold the variable in a field of the record.
\begin{lstlisting}
Record contract := Contract {
    d : set ident ;
    A : assertion d ;
    G : assertion d ;
}.
\end{lstlisting}

But the problem occurs when defining operators on contracts.
We need contracts to be defined on the same variables.
If the contracts are defined on different variables, we don't have any definition of the operator.
This means we should have a partial function, returning an option type, namely, it returns the result when it exists, and a default value when it does not.
However, working with partial functions means we have to always verify that the result exists when proving theories about it.
By parameterizing the type of contract we limit the use of the operator to only contracts defined on the same variables.
This means the operator is a total function which is easier to work with, especially in proof activity.

  \subsection{Extending Assertion on Another Set of Variables}

We first defined the extension of assertion by using the union of the variables it's defined on, and another set of variables.
\begin{lstlisting}
Definition extend_assertion {d1 d2 : set ident} (a1 : assertion d1) :
    assertion (d1 $\cup$ d2) := ...
\end{lstlisting}

This definition adds a lot of problems when composing contracts defined on variable sets that are equal but constructed differently.
This leads to an assertion defined on \lstinline{d1 $\cup$ d2}, and another on \lstinline{d2 $\cup$ d1}.
The two sets are equal, thanks to the union being commutative, but for instance, the types \lstinline{d1 $\cup$ d2} and \lstinline{d2 $\cup$ d1} are not the same in Coq.
Since we parameterize the contracts type with their variables of definition, the types \lstinline{contract (d1 $\cup$ d2)} and \lstinline{contract (d2 $\cup$ d1)} are also different.
Our composition operator requires the two contracts to be of the same type, this means it is not possible to compose contracts that are not of the same types.
Composing these contracts is impossible, we need to change their definitions.

One solution would be to define another extension function:
\begin{lstlisting}
Definition extend_assertion_l {d1 d2 : set ident} (a2 : assertion d2) :
    assertion (d1 $\cup$ d2) := ...
\end{lstlisting}
But it seems quite inelegant, every function needs to be defined two times.
Our solution is to use another set of variables as the final set.
\begin{lstlisting}
Definition extend_assertion {H : d1 $\subseteq$ d2} (a1 : assertion d1) :
    assertion d2 := ...
\end{lstlisting}
This removes the problem altogether but changes the way we have to think about the extension of assertion.

\section{Conclusion}

In this paper, we presented a formalization of the set-theoretical assume/guarantee contracts in the proof assistant Coq, and showed how to instantiate it with a given logic.
To the best of our knowledge, it is the first mechanized formalization of a theory of
assume/guarantee contracts for system design.
The formalization gives us the assurance that the notion of assume/guarantee contract is a correct instance of the meta-theory of contract.
We also gave a set-theoretic definition of extension and elimination of variables in a contract, which was not defined in the original works.
Finally, we demonstrated how to construct contracts in a simple propositional logic and proved the refinement, conjunction, and composition rules correct.
The complete implementation of our formalization in Coq is available \href{https://github.com/merce-fra/SKT-VerifedContractTheory}{\textbf{here}}.%
\footnote{\href{https://github.com/merce-fra/SKT-VerifedContractTheory}{https://github.com/merce-fra/SKT-VerifedContractTheory}}

Ideally, the theory of contracts should help engineers struggle with the specification of contracts
during system design.
In this aim, having a tool to detect contradictory contracts early could prove useful.
For now, the composition operator does not certify that the resulting contract is compatible with any environment nor implementable by a component.
This may lead to a problem in the design process should the contradictory contracts be composed and incorrect contracts be defined, leading to an unimplementable contract.
We can detect that a contract is unimplementable but it could be useful for a more efficient process to be able to detect contradictory contracts before composing them.

While the formalization of the contract theory we made (in the proof assistant Coq) may be of cumbersome use for realistically scaled design systems, this work was not intended to provide a usable contract theory applicable to any logic.
Each logic features specific design choices that hint at the proper way contracts should be combined with them.
However, our formalization demonstrates that different contract definitions fit the same global (meta) theory.
Our aim is to prove that their definitions of refinement, composition, and conjunction are equivalent.
The first step of this future work would be to implement a hybrid logic such as differential dynamic logic or the duration calculus into the Coq proof assistant, then formalize their contract theory.
Finally, by instantiating our contract theory, we could show the equivalence of their contracts, or pinpoint their differences if they are not equivalent.

\bibliographystyle{splncs04}
\bibliography{biblio}

\begin{thebibliography}{10}
\providecommand{\url}[1]{\texttt{#1}}
\providecommand{\urlprefix}{URL }
\providecommand{\doi}[1]{https://doi.org/#1}

\bibitem{abadi_composing_1993}
Abadi, M., Lamport, L.: Composing specifications. ACM Transactions on
  Programming Languages and Systems  \textbf{15}(1),  73--132 (Jan 1993).
  \doi{10.1145/151646.151649}

\bibitem{arechiga_specifying_2019}
Ar{\'e}chiga, N.: Specifying {{Safety}} of {{Autonomous Vehicles}} in {{Signal
  Temporal Logic}}. In: 2019 {{IEEE Intelligent Vehicles Symposium}} ({{IV}}).
  pp. 58--63 (Jun 2019). \doi{10.1109/IVS.2019.8813875}

\bibitem{benveniste_multiple_2008}
Benveniste, A., Caillaud, B., Ferrari, A., Mangeruca, L., Passerone, R.,
  Sofronis, C.: Multiple {{Viewpoint Contract}}-{{Based Specification}} and
  {{Design}}. In: {de Boer}, F.S., Bonsangue, M.M., Graf, S., {de Roever}, W.P.
  (eds.) Formal {{Methods}} for {{Components}} and {{Objects}}. pp. 200--225.
  Lecture {{Notes}} in {{Computer Science}}, {Springer}, {Berlin, Heidelberg}
  (2008). \doi{10.1007/978-3-540-92188-2_9}

\bibitem{benveniste_contracts_2015}
Benveniste, A., Caillaud, B., Nickovic, D., Passerone, R., Raclet, J.B.,
  Reinkemeier, P., {Sangiovanni-Vincentelli}, A., Damm, W., Henzinger, T.,
  Larsen, K.G.: Contracts for {{Systems Design}}: {{Theory}}. Report, {INRIA}
  (Jul 2015)

\bibitem{chaochen_formal_1996}
Chaochen, Z., Ji, W., Ravn, A.P.: A formal description of hybrid systems. In:
  Alur, R., Henzinger, T.A., Sontag, E.D. (eds.) Hybrid {{Systems III}}. pp.
  511--530. Lecture {{Notes}} in {{Computer Science}}, {Springer}, {Berlin,
  Heidelberg} (1996). \doi{10.1007/BFb0020972}

\bibitem{chaochen_extended_1993}
Chaochen, Z., Ravn, A.P., Hansen, M.R.: An extended duration calculus for
  hybrid real-time systems. In: Grossman, R.L., Nerode, A., Ravn, A.P.,
  Rischel, H. (eds.) Hybrid {{Systems}}. pp. 36--59. Lecture {{Notes}} in
  {{Computer Science}}, {Springer}, {Berlin, Heidelberg} (1993).
  \doi{10.1007/3-540-57318-6_23}

\bibitem{chen_mars_2017}
Chen, M., Han, X., Tang, T., Wang, S., Yang, M., Zhan, N., Zhao, H., Zou, L.:
  {{MARS}}: {{A Toolchain}} for {{Modelling}}, {{Analysis}} and
  {{Verification}} of {{Hybrid Systems}}. In: Hinchey, M., Bowen, J.P.,
  Olderog, E.R. (eds.) Provably {{Correct Systems}}, pp. 39--58. {{NASA
  Monographs}} in {{Systems}} and {{Software Engineering}}, {Springer
  International Publishing}, {Cham} (2017)

\bibitem{dabney_mastering_2003}
Dabney, J.B., Harman, T.L.: Mastering {{Simulink}}. {Pearson}, {Upper Saddle
  River, N.J} (Nov 2003)

\bibitem{foster_unifying_2019}
Foster, S., Cavalcanti, A., Canham, S., Woodcock, J., Zeyda, F.: Unifying
  theories of reactive design contracts. Theoretical Computer Science
  \textbf{802},  105--140 (Jan 2020). \doi{10.1016/j.tcs.2019.09.017}

\bibitem{fulton_keymaera_2015}
Fulton, N., Mitsch, S., Quesel, J.D., V{\"o}lp, M., Platzer, A.:
  {{KeYmaera~X}}: {{An Axiomatic Tactical Theorem Prover}} for {{Hybrid
  Systems}}. In: Felty, A.P., Middeldorp, A. (eds.) Automated {{Deduction}} -
  {{CADE}}-25. pp. 527--538. Lecture {{Notes}} in {{Computer Science}},
  {Springer International Publishing}, {Cham} (2015).
  \doi{10.1007/978-3-319-21401-6_36}

\bibitem{graf_building_2018}
Graf, S., Quinton, S., Girault, A., G{\"o}ssler, G.: Building {{Correct
  Cyber}}-{{Physical Systems}}: {{Why We Need}} a {{Multiview Contract
  Theory}}. In: Howar, F., Barnat, J. (eds.) Formal {{Methods}} for
  {{Industrial Critical Systems}}. pp. 19--31. Lecture {{Notes}} in {{Computer
  Science}}, {Springer International Publishing}, {Cham} (2018).
  \doi{10.1007/978-3-030-00244-2_2}

\bibitem{liebrenz_deductive_2018}
Liebrenz, T., Herber, P., Glesner, S.: Deductive {{Verification}} of {{Hybrid
  Control Systems Modeled}} in {{Simulink}} with {{KeYmaera X}}. In: Sun, J.,
  Sun, M. (eds.) Formal {{Methods}} and {{Software Engineering}}. pp. 89--105.
  Lecture {{Notes}} in {{Computer Science}}, {Springer International
  Publishing}, {Cham} (2018). \doi{10.1007/978-3-030-02450-5_6}

\bibitem{lunel_parallel_2019}
Lunel, S., Mitsch, S., Boyer, B., Talpin, J.P.: Parallel {{Composition}} and
  {{Modular Verification}} of {{Computer Controlled Systems}} in {{Differential
  Dynamic Logic}}. In: ter Beek, M.H., McIver, A., Oliveira, J.N. (eds.) Formal
  {{Methods}} - {{The Next}} 30 {{Years}} - {{Third World Congress}}, {{FM}}
  2019, {{Porto}}, {{Portugal}}, {{October}} 7-11, 2019, {{Proceedings}}.
  Lecture {{Notes}} in {{Computer Science}}, vol. 11800, pp. 354--370.
  {Springer} (2019). \doi{10.1007/978-3-030-30942-8_22}

\bibitem{meyer_applying_1992}
Meyer, B.: Applying 'design by contract'. Computer  \textbf{25}(10),  40--51
  (Oct 1992). \doi{10.1109/2.161279}

\bibitem{muller_tactical_2018}
M{\"u}ller, A., Mitsch, S., Retschitzegger, W., Schwinger, W., Platzer, A.:
  Tactical contract composition for hybrid system component verification.
  International Journal on Software Tools for Technology Transfer
  \textbf{20}(6),  615--643 (Nov 2018). \doi{10.1007/s10009-018-0502-9}

\bibitem{nipkow_isabellehol_2002}
Nipkow, T., Paulson, L.C., Wenzel, M.: Isabelle/{{HOL}}: {{A Proof Assistant}}
  for {{Higher}}-{{Order Logic}}. Lecture {{Notes}} in {{Computer Science}},
  {{Lect}}.{{Notes Computer}}. {{Tutorial}}, {Springer-Verlag}, {Berlin
  Heidelberg} (2002). \doi{10.1007/3-540-45949-9}

\bibitem{nuzzo_compositional_2015}
Nuzzo, P.: Compositional {{Design}} of {{Cyber}}-{{Physical Systems Using
  Contracts}}. Ph.D. thesis, UC Berkeley (2015)

\bibitem{platzer_differential_2008}
Platzer, A.: Differential {{Dynamic Logic}} for {{Hybrid Systems}}. Journal of
  Automated Reasoning  \textbf{41}(2),  143--189 (Aug 2008).
  \doi{10.1007/s10817-008-9103-8}

\bibitem{sangiovanni-vincentelli_taming_2012}
{Sangiovanni-Vincentelli}, A., Damm, W., Passerone, R.: Taming {{Dr}}.
  {{Frankenstein}}: {{Contract}}-{{Based Design}} for {{Cyber}}-{{Physical
  Systems}}*. European Journal of Control  \textbf{18}(3),  217--238 (Jan
  2012). \doi{10.3166/ejc.18.217-238}

\bibitem{the_coq_development_team_coq_2018}
Team, T.C.D.: The {{Coq Proof Assistant}}, version 8.7.2. Zenodo (Feb 2018).
  \doi{10.5281/zenodo.1174360}

\bibitem{wang_assumeguarantee_2012}
Wang, S., Zhan, N., Guelev, D.: An {{Assume}}/{{Guarantee Based Compositional
  Calculus}} for {{Hybrid CSP}}. In: Agrawal, M., Cooper, S.B., Li, A. (eds.)
  Theory and {{Applications}} of {{Models}} of {{Computation}}. pp. 72--83.
  Lecture {{Notes}} in {{Computer Science}}, {Springer}, {Berlin, Heidelberg}
  (2012). \doi{10.1007/978-3-642-29952-0_13}

\bibitem{wang_improved_2015}
Wang, S., Zhan, N., Zou, L.: An {{Improved HHL Prover}}: {{An Interactive
  Theorem Prover}} for {{Hybrid Systems}}. In: Butler, M., Conchon, S.,
  Za{\"i}di, F. (eds.) Formal {{Methods}} and {{Software Engineering}}. pp.
  382--399. Lecture {{Notes}} in {{Computer Science}}, {Springer International
  Publishing}, {Cham} (2015). \doi{10.1007/978-3-319-25423-4_25}

\end{thebibliography}

\end{document}